\documentclass[oldversion]{aa} % double column 
\usepackage{graphicx}
\usepackage{aalongtable}
\usepackage{txfonts}
\usepackage{rotating}
\usepackage{lscape}
\newcommand{\gsim}{\;\lower.6ex\hbox{$\sim$}\kern-7.75pt\raise.65ex\hbox{$>$}\;}
\newcommand{\lsim}{\;\lower.6ex\hbox{$\sim$}\kern-7.75pt\raise.65ex\hbox{$<$}\;}

\begin{document}
\title{A comparison of abundance analyses of first generation stars in multiple
populations in 47~Tuc and NGC~3201
 }

\author{
Eugenio Carretta\inst{1}
\and
Angela Bragaglia\inst{1}
}

\authorrunning{Carretta and Bragaglia}
\titlerunning{Comparison of abundance analyses in FG stars}

\offprints{E. Carretta, eugenio.carretta@inaf.it}

\institute{
INAF-Osservatorio di Astrofisica e Scienza dello Spazio di Bologna, Via P. Gobetti
 93/3, I-40129 Bologna, Italy}

\date{}

\abstract{The distinction of the stellar content in globular clusters (GCs) in
multiple stellar populations characterized by different amounts of
proton-capture elements has been well assessed since a long time. On the other
hand, the existence of noticeable variations in metallicity among GC stars is
still debated. In particular, recent spectroscopic analyses claimed the presence
of a small variation in metallicity, $\sim 0.1$ dex, for the first generation
(FG) stars in NGC~3201 and NGC~104. However, in both cases the claim is not
robust because of the internal error of 0.1 dex associated to the  [Fe/H]
values. To verify the reality of a metallicity variation we compared the two
analyses, performed by the same authors with identical methodology. We found
trends of metallicity as a function of the spectroscopically derived effective
temperatures. However they are in opposite directions; in NGC~3201 cooler (and
brighter) stars have higher [Fe/H] values, whereas in 47~Tuc they show lower
metallicities. The trend is not statistically significant in the former case,
but it is in the latter. The dependence of  metallicity on the luminosity along
the red giant branch seems to indicate problems in the abundance analysis for 47
Tuc. Finally, effective temperature do not show a significant variation as a
function of the colour spread along the HST pseudo-colour map, which we should
instead observe were the trends temperature-metallicity a real effects of
intrinsic scatter in iron. According to this comparison, we conclude that with 
these analyses, and the associated spurious trends, the issue of metallicity
variations in FG stars is hardly settled.
}

\maketitle

\section{Introduction}

Among the many unsolved issues concerning multiple stellar populations (MPs)
in globular clusters (GCs), one of the most surprising and recent is 
the possible existence of a metallicity spread in first generation
(FG) stars. These stars are recognised as the original population formed in a
primary burst of star formation at the beginning of the GC lifetime. Polluters
within this first generation provided the matter processed by proton-capture
reactions in H-burning at high temperature. The mixing of this processed matter
with pristine gas explains the composition of the polluted stars of the second
generation (SG), as well as the correlation and anti-correlation of light
elements bearing clear traces of proton-capture reactions (He, C, N, O, Mg, Al, 
sometime K, Ca, Sc, see e.g. the extensive reviews. by Gratton et al. 2004,
2012,  2019, Bastian and Lardo 2018).

The recent claim of iron spreads among FG stars is surprising in light of the fact
that most of candidate polluters proposed to enrich the pristine proto-cluster
environment with processed ejecta are not able to produce metallicity variations.
Asymptotic giant branch stars (AGB: Ventura et al. 2001), fast rotating massive
stars (Decressin et al. 2007), interacting massive binaries (de Mink et al. 2009),
supermassive stars (Denissenkov and Hartwick 2014) may reproduce only the
relationship between light elements, although with some quantitative problems. 

To also alter the cluster content of iron (and heavier elements), supernova (SN)
nucleosynthesis must necessarily be involved. Furthermore, all the above
candidates are massive stellar objects, implying short evolving times and a time
range for their action restricted to a maximum of a few tens of million years 
for the slowest evolving among them (AGB stars). However, the nearly homogeneity
of heavy element in the majority of GCs (better than 12\% or less than 0.05 dex,
Carretta et al. 2009a) implies that ejecta from SNe should not be retained
in GCs contributing to the secondary star formation.

A few GCs in the Milky Way are known to host intracluster metallicity
variations. The amount of star to star differences in [Fe/H]\footnote{We adopt
the usual spectroscopic notation, $i.e.$  [X]= log(X)$_{\rm star} -$
log(X)$_\odot$ for any abundance quantity X, and  log $\epsilon$(X) = log
(N$_{\rm X}$/N$_{\rm H}$) + 12.0 for absolute number density abundances.} is
varying. On one hand, GCs such as NGC~1851 show a spread just barely larger than
the observational errors (Carretta et al. 2011). At the other extreme, the most
massive GC in the Galaxy, NGC~5139 ($\omega$ Cen) shows
clear evidence of star to star variations for all elements (see e.g. Gratton
et al. 2004, Johnson and Pilachowski 2010). The possibility that this GC was the
nucleus of a dissolved nucleated dwarf galaxy was speculated upon by many
studies (e.g. Zinnecker et al. 1988, Freeman 1993). The second most massive
Galactic GC, M~54, may be also the nucleus of another dwarf, Sagittarius.
Analogies and differences between these two massive GCs can be attributed to a
common dynamical evolution, but observed at different stages (Carretta et al.
2010a). To explain the few  iron complex GCs, peculiar dynamical conditions for
their formation in dwarf galaxy satellites (Bekki and Tsujimoto 2016) and
mechanisms of inhomogeneous chemical evolution following cloud-cloud collisions
(e.g. Tsujimoto and Shigeyama 2003) have been proposed.

Nevertheless, in most GCs the level of metallicity variations rarely exceed the
typical internal errors of precise abundance analysis, that is about 0.05 dex.
So why are we talking of metallicity spread in FG stars of objects that are
generally found to be very homogeneous in iron on the basis of spectroscopic
analysis?

Using HST photometry with appropriate filters sampling the spectral regions
where molecular features of CNO elements are preferentially located, Milone et
al. (2017) arranged the stars in 57 GCs in a pseudo-colour map (PCM). On the
PCMs stars are mostly located in two regions, corresponding to the groups of
stars having primordial composition and chemistry altered by proton-capture
processes, hereafter $RG1$ and $RG2$, respectively (see the simplified notation in
Tab.~\ref{t:notation}, borrowed from Carretta and Bragaglia 2024). 

Milone and collaborators estimated that the $RG1$ in PCMs were too much extended
with respect to observational errors obtained by Monte Carlo simulations. Since
the horizontal coordinate $\Delta col$ is related to shifts in effective
temperatures of stars, sampled by a colour with long baseline in wavelength,
they claimed that star to star variations in metallicity or in the He content
could explain the extended $RG1$ sequence. Since FG stars are by definition
objects where the He mass fraction Y is untouched since Big Bang
nucleosynthesis, the consequent inference was that variations among them
should be attributed to spread in [Fe/H] (see also Legnardi et al. 2022).

In a differential high resolution abundance analysis of five FG stars in NGC~2808 
Lardo et al. (2023) found a correlation between abundances and position along
$RG1$, with lower content in Si, Ca, Ti~{\sc ii} and Ni for stars with more
negative values of $\Delta col$,  in agreement with the above inference from
photometry. For [Fe/H] they derived  a range equal to  $0.25\pm 0.06$ dex. This
result is conflicting with the  analysis of a large sample of stars in NGC~2808 by
Carretta et al. (2006) who set a limit $\lesssim 0.05$ dex to the maximum iron
spread in this GC, a result later confirmed by Carretta (2015). Moreover, Carretta
et al. (2006) found that the average iron abundance in NGC~2808 was increasing
going from stars with primordial composition to stars increasingly enriched in
products from proton-capture reactions. Since He is the main outcome of the
required H-burning, this was in qualitative agreement with an increased strength
of metallic lines due to the increase in He content (B\"ohm-Vitense 1979).
Unfortunately, the statistical significance of this result was not very high, due
to the attached observational errors. On the other hand,  even if the 
correlations are clear in Lardo et al. (2023), none has a high statistical
significance, due to the very limited number of stars.

To delve more in depth on the reality of metallicity spreads in FG stars, we test
here the above inference by comparing the abundance analyses of FG stars in two
GC, namely NGC~3201 (Marino et al. 2019: M19) and NGC~104 (47 Tuc; Marino et al.
2023, hereinafter M23). 

The idea is simple. If the metallicity spread is an intrinsic property of FG
stars in GCs, we should see similar patterns standing out in both GCs under
scrutiny, regardless of their difference in global metallicity 
([Fe/H]$=-0.768$ dex for 47~Tuc and [Fe/H]$=-1.512$ dex for NGC~3201, Carretta
et al. 2009a).
The advantage is that both analyses were performed by the same group, with the
same instrument and very similar (if not identical) techniques for the abundances
obtained from high resolution spectra. We thus aim to highlight whether the
findings are consistent and depict a coherent scenario or else there are
discrepancies that weaken the conclusion about the metallicity spread in FG stars.

In this paper we first stress the implications of a claimed metallicity spread
when the internal observational errors are of the same magnitude (Section 2).
Then we compare trends of the derived metallicities in the two GCs with
parameters and $\Delta col$ values, to ascertain how and how well the variations
in effective  temperatures T$_{\rm eff}$ are matched by shift in metallicity
along the PCM (Section 3). A discussion and final considerations on the observed
trend in the two GCs under scrutiny are finally given in Section 4.

\begin{table}
\centering
\caption[]{Notation adopted in the present paper}
\begin{tabular}{l}
\hline

$col = mag_{F275W}-mag_{F814W}$ \\

$col3 = (mag_{F275W}-mag_{F336W}) - (mag_{F336W}-mag_{F438W})$ \\

$xr = col$ of the fiducial red line \\ 

$xb = col$ of the fiducial blue line \\ 

$yr = col3$ of the fiducial red line \\ 

$yb = col3$ of the fiducial blue line \\ 

 $\Delta col = W_{col} ( (col -xr)/(xr-xb) )$  \\
 
 $\Delta col3 = W_{col3} ( (col3 -yr)/(yr-yb) ) $  
(see Section 2) \\

 $W_{col}$ and $W_{col3}$ = widths of the RGB in $col$ and $col3$\\

PCM = pseudo-colour map = $\Delta col$ in abscissa and $\Delta col3$ in ordinate \\

$RG1$ = region in the PCM populated by FG stars only\\

$RG2$ = region in the PCM populated by SG (or non-FG) stars\\

\hline
\end{tabular}
\label{t:notation}
\end{table}

\begin{figure*}
\centering
\includegraphics[scale=0.30]{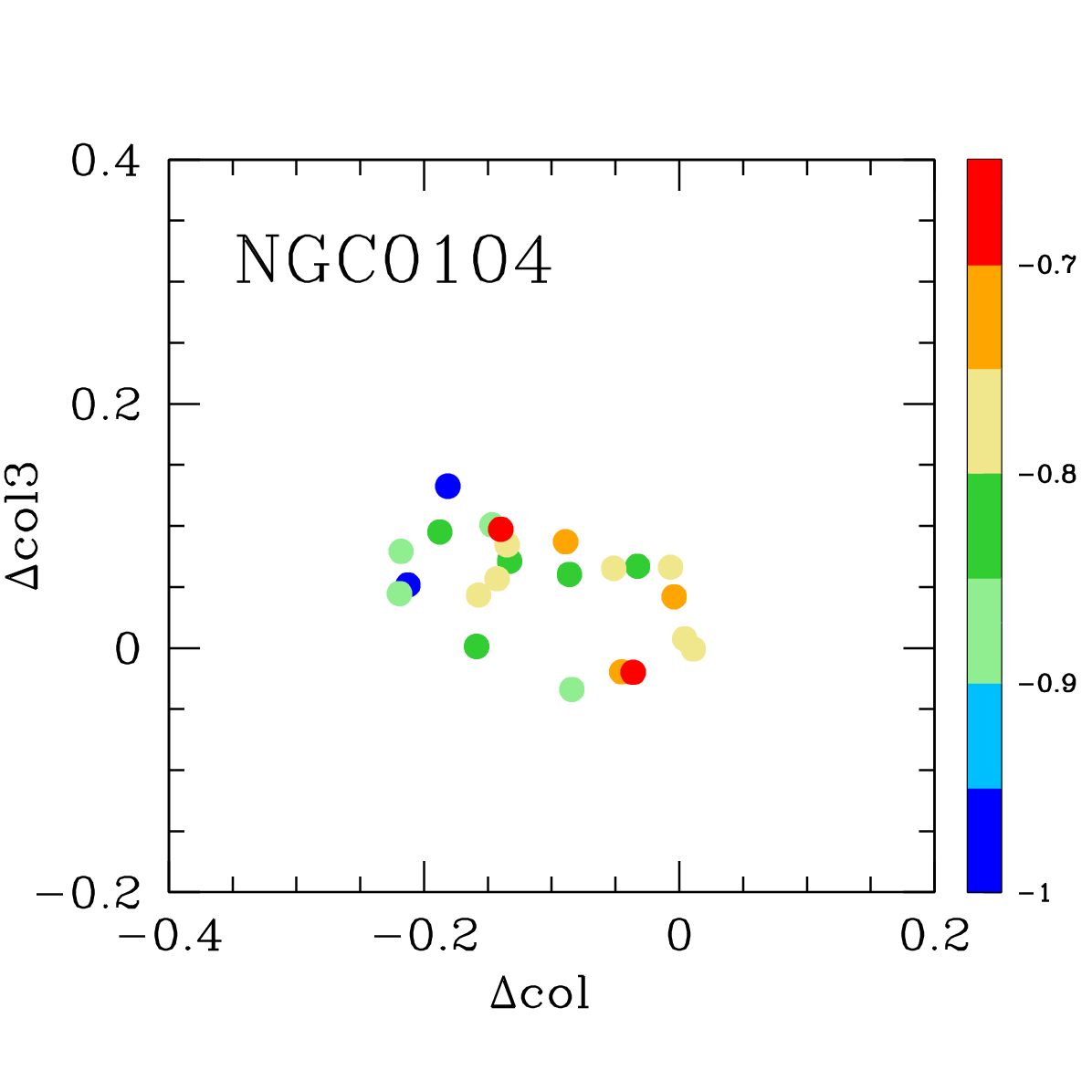}\includegraphics[scale=0.30]{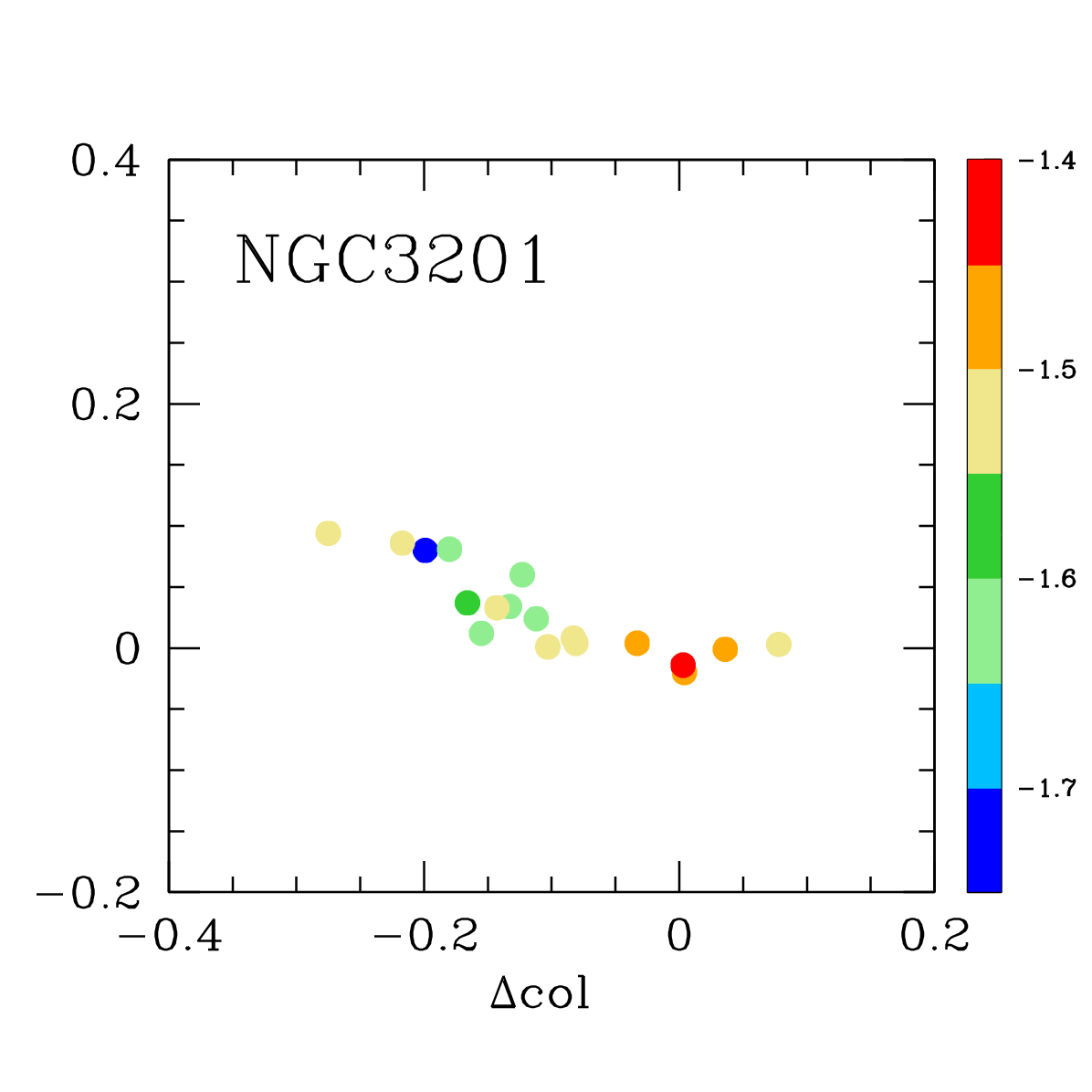}
\includegraphics[scale=0.30]{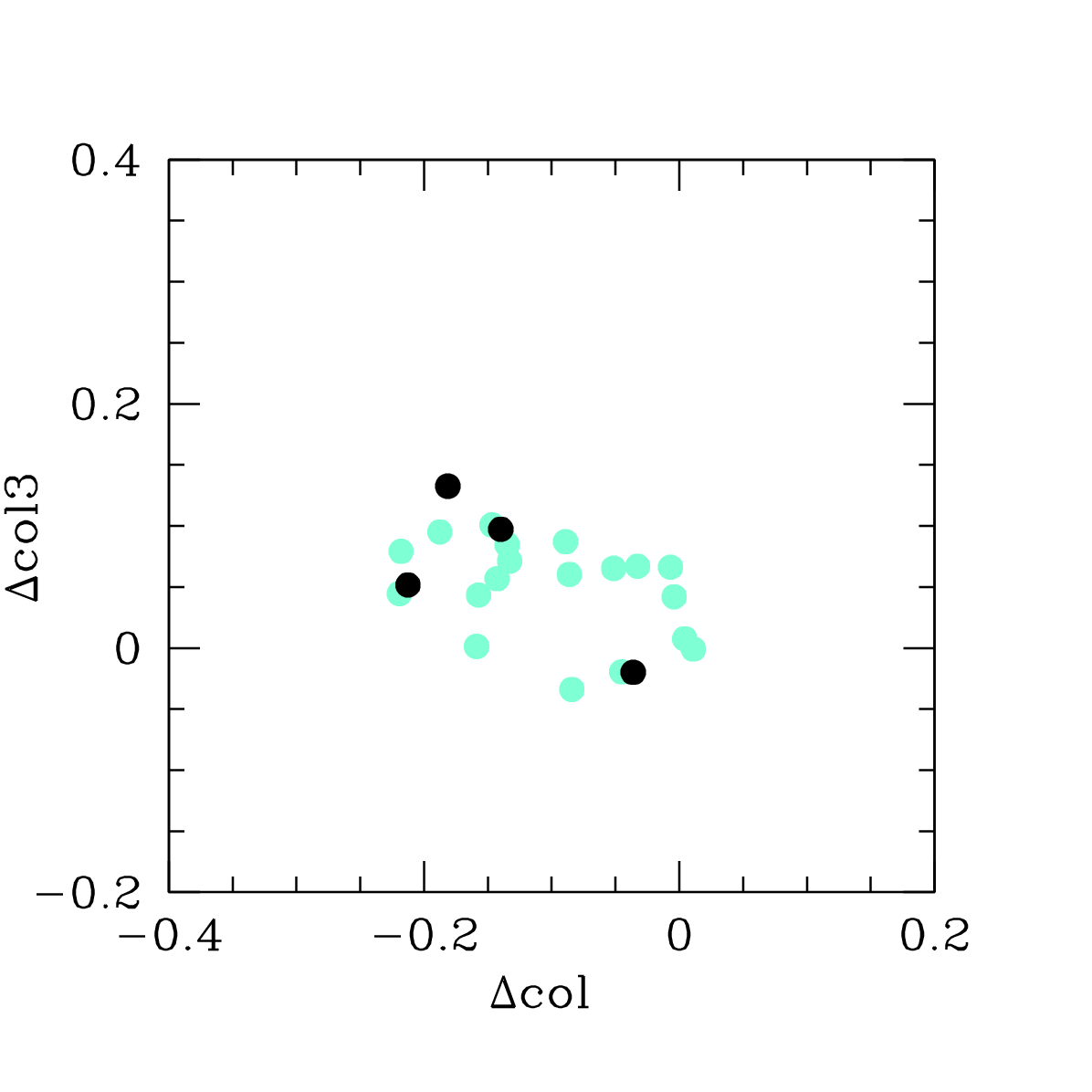}\includegraphics[scale=0.30]{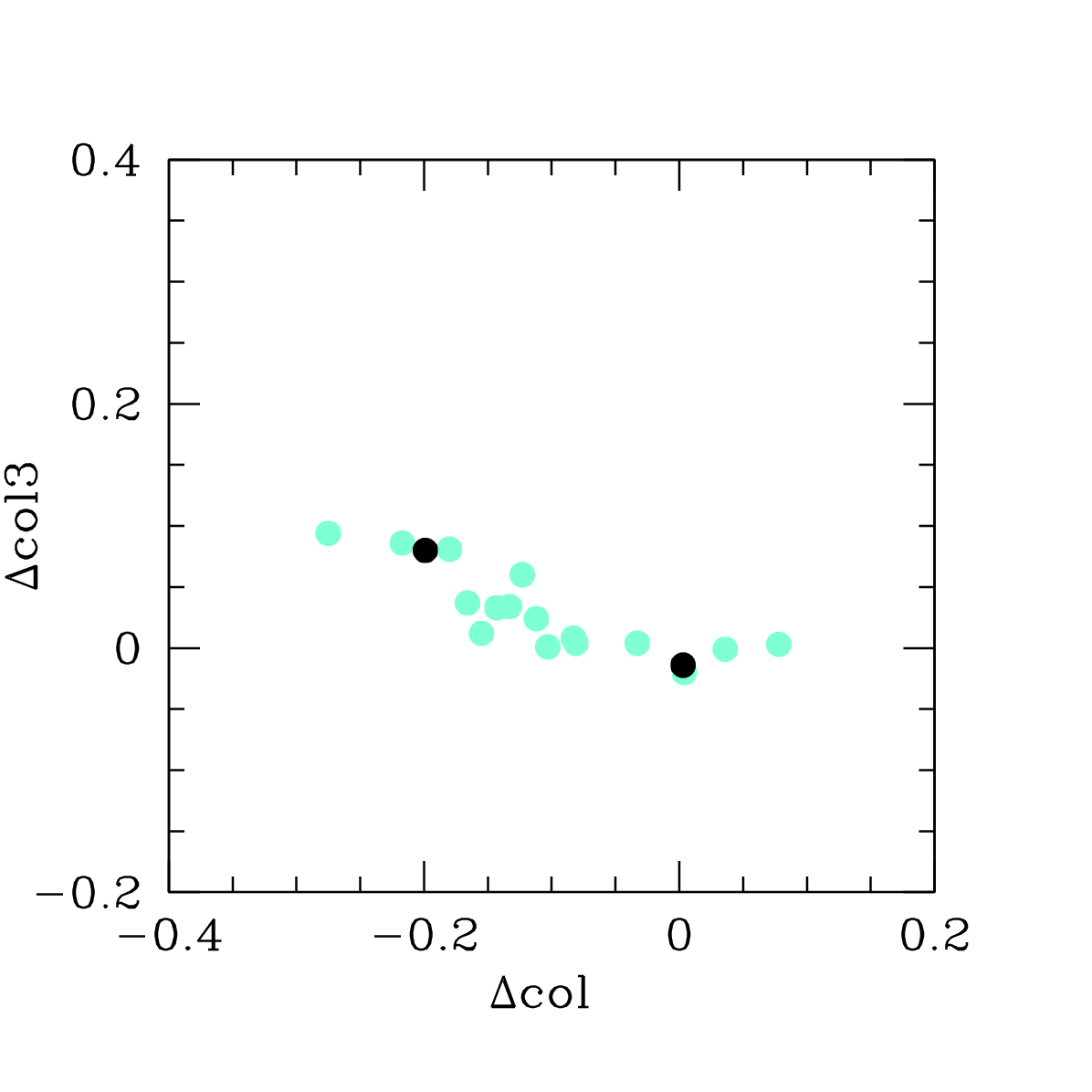}
\caption{Pseudo-colour maps of 47~Tuc (upper left panel) and NGC~3201 
(upper right panel) from the only published data in M23 and M19, respectively.
Stars are colour-coded according to their metallicity. In the lower panels, the
aqua colour is used for all stars having [Fe/H] values within $\pm 0.10$ dex (the
internal error associated to the metallicity) from the average metallicity of FG
stars in each cluster, whereas the outliers of this range are marked in black.} 
\label{f:fig0}
\end{figure*}

\section{Intrinsic spreads with an internal error of 0.1 dex in [Fe/H]}

The data compared in the present paper are the abundances of 18 FG stars in
NGC~3201 and 23 FG stars in 47~Tuc derived by M19 and M23, respectively, using
high resolution spectra from the FLAMES-UVES (Ultraviolet and Visual Echelle
Spectrograph) instrument on the ESO-VLT UT2 telescope. In both cases stellar
parameters (T$_{\rm eff}$, $\log g$, [A/H], $v_t$) were derived using a fully
spectroscopic analysis. 

Complete PCMs from Milone et al. (2017) are not available because 
they were never published. However, the $\Delta col$ and $\Delta col3$
values for the FG stars analysed are given in M19 and M23, allowing to evaluate
the  displacements of stars along the $RG1$.

Excluding three binary candidates, M19 found a small metallicity range of  $\sim
0.10$ dex among the other 15 FG giants in NGC~3201. This same amount of
variations was also found in the sample of FG stars in 47~Tuc (M23), where two
candidate binaries were instead left in the sample. However, both abundance
analyses came with an attached internal error of 0.10 dex in metallicity (M19
and M23). The same amount found for errors and claimed spread in [Fe/H] 
raises doubts on the reality of the spread.

The concept is visualised in Fig.~\ref{f:fig0}. In the upper panels we plot the
sample of FG stars in 47~Tuc (left upper panel) and in NGC~3201 (right upper 
panel) where the giants are colour-coded according to the metallicities derived
in M19 and M23. In the lower panels, we plot the
same stars, but this time assigning the same colour to all stars within 
$\pm 0.1$dex from the average [Fe/H] value of the GCs. The result is that there
seems to be no clear relation between $\Delta col$ and metallicity spread. Apart
from a very few outliers, the majority of stars share a common metallicity,
within the associated internal error of 0.1 dex.

Another occurrence immediately evident is that the $RG1$ sequence in NGC~3201 is
about 0.15 mag more extended than the one in 47~Tuc. Since the observational
error (estimated from the figures in Milone et al. 2017) is about 0.05 mag in
$\Delta col$, this is a 3$\sigma$ difference. However, for both GCs the derived
metallicity spread is the same, namely 0.1 dex. 

A more quantitative estimate of the intrinsic spread in presence of a given
internal error is obtained by the approach suggested in Mucciarelli et al.
(2012). The method searches in a grid of the ([Fe/H], $\sigma$) space, where
$\sigma$ is the intrinsic spread, the couple of parameters that maximises a
maximum likelihood function. This function is defined accounting for the number
of stars in the sample and the associated error.  Employing this approach with
the datasets of FG stars in the two GCs by Marino et al. we found null intrinsic
spreads, that is $\sigma = 0.00 \pm 0.031$ dex and  $\sigma = 0.00 \pm 0.032$
dex for 47~Tuc and NGC~3201, respectively, when an internal error of 0.1 dex is
considered.

\section{Comparison of abundance analyses in 47~Tuc and NGC~3201}

The location of stars along $RG1$ should be a direct consequence of the position
of stars on the colour-magnitude diagram (CMD). In the plane $m_{F814W}$ versus
$col$, more metal-poor stars lie on the blue side of the RGB, whereas more
metal-rich stars populate the red side. This is a general property of stellar
models (e.g. Salaris et al. 2002). For a given absolute brightness, the RGB
colour (i.e. the stellar T$_{\rm eff}$) is strongly affected by the value of
metallicity. Even when differences in colour from red and fiducial lines are
combined in a PCM stars with lower metallicity should be found at bluer values
of $\Delta col$ along $RG1$. In short, this is the basis of the claims for a
metallicity spread among FG stars (M19, Lardo et al. 2023, M23).

\begin{figure*}
\centering
\includegraphics[scale=0.30]{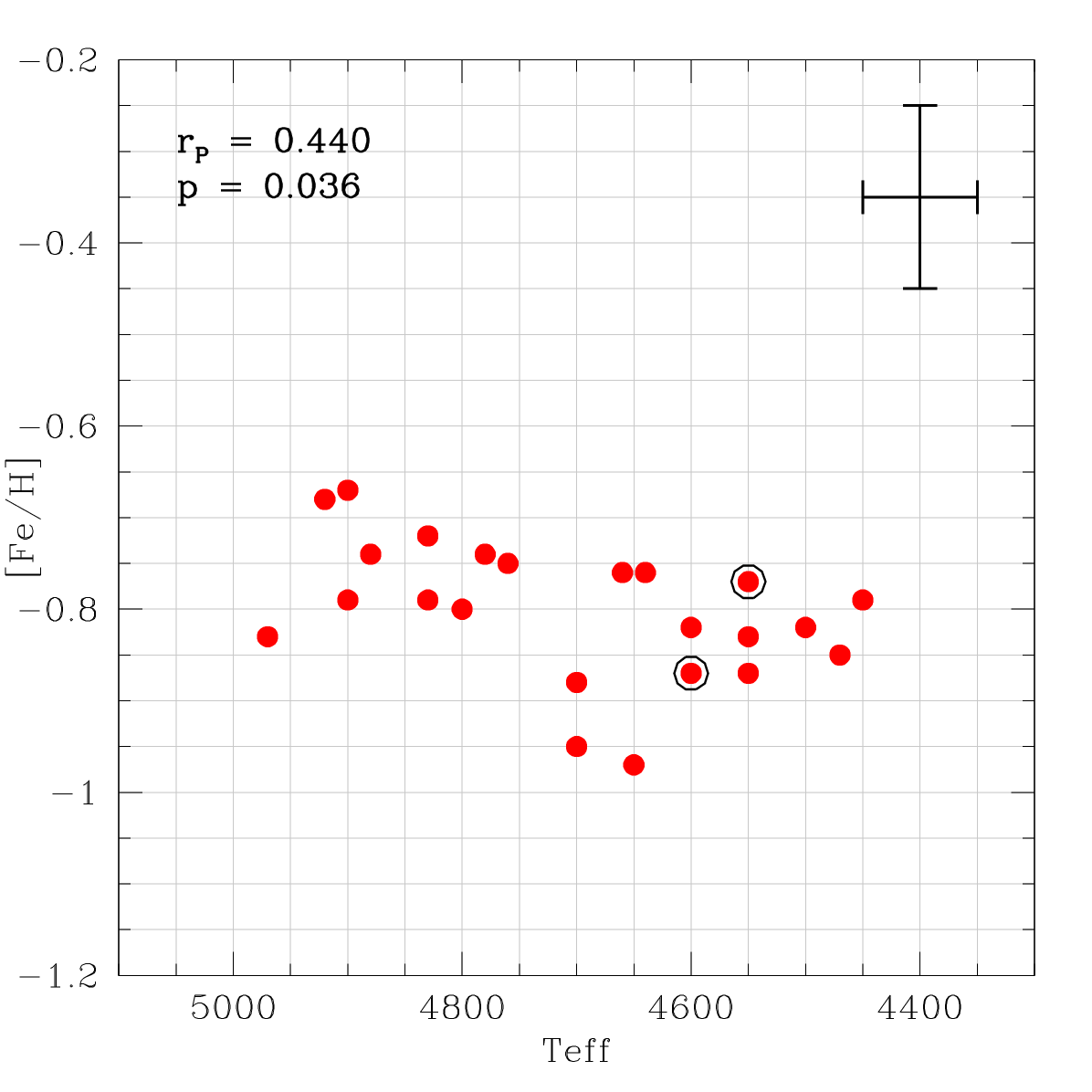}\includegraphics[scale=0.30]{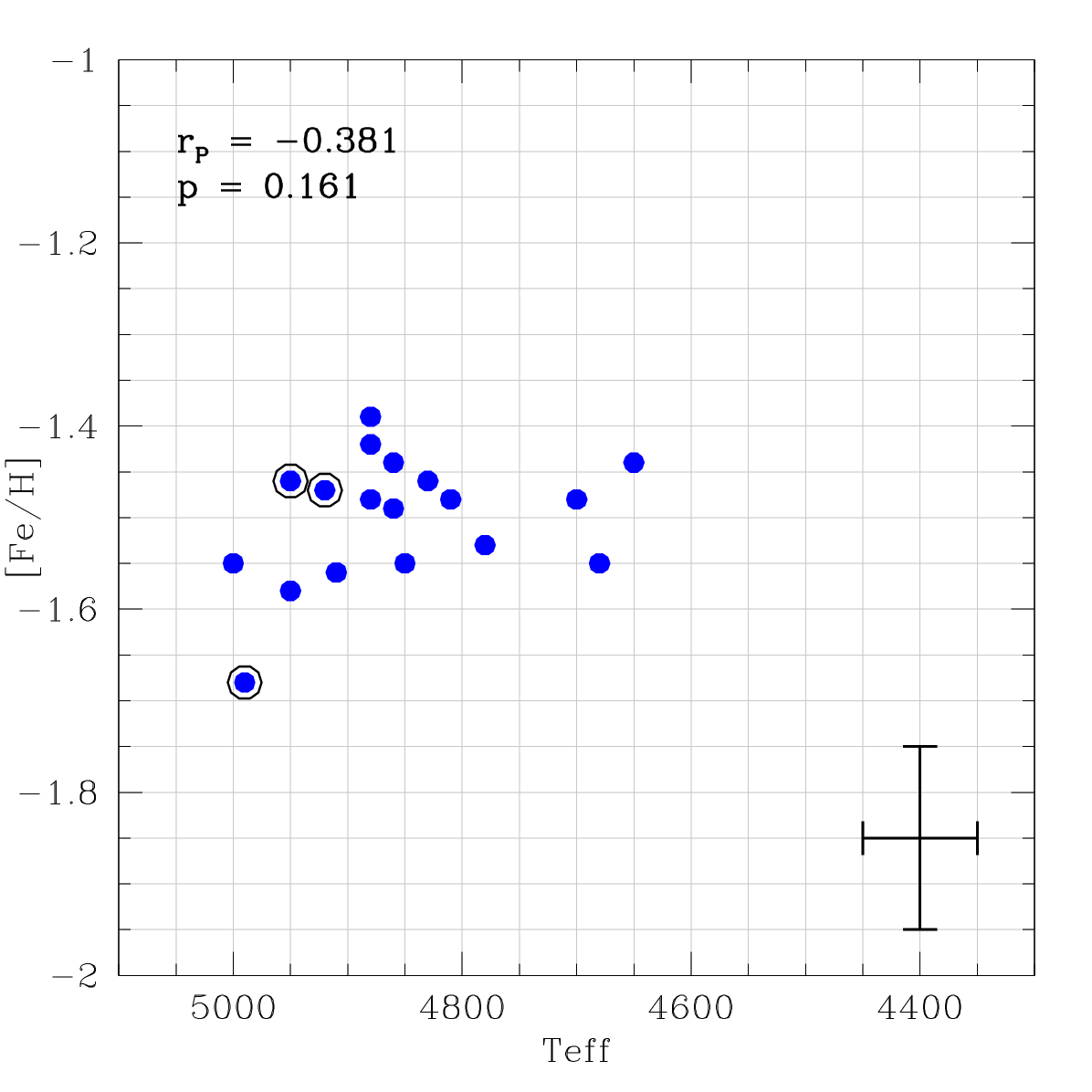}
\caption{Relations between T$_{\rm eff}$ and (derived) metallicity [Fe/H] in FG
stars of 47~Tuc (left panel, red points) and NGC~3201 (right panel, blue points)
from M23 and M19, respectively. In each panel the Pearson's coefficient for the
linear regression and the two-tail probability that the true correlation is zero
are listed. Black circled points are candidate binary stars (here and in the 
following figures). Error-bars indicate typical errors taken from M19 and
M23.}
\label{f:feteff}
\end{figure*}

In Fig.~\ref{f:feteff} we show the metallicity derived for FG stars in 47~Tuc
(left panel, from M23) and in NGC~3201 (right panel, from M19) as a function of
the effective temperature of stars. The superimposed grid helps to visualise the
gradients existing in the data. In 47~Tuc, cooler stars are found to be more 
metal-poor than stars with higher temperatures. This trend is opposite to the
expectations from the colour-metallicity dependence of RGB models, hence it is
likely to be a spurious occurrence due to the abundance analysis.

To test the statistical significance of this trend, we computed a linear
regression between [Fe/H] and T$_{\rm eff}$. In the left panel of
Fig.~\ref{f:feteff} we list the Pearson's correlation coefficient and the
two-tail probability that tests the null hypothesis that the observed values
come from a population in which the true correlation is zero. We obtained a
probability $p=0.036$ (23 stars), hence the observed correlation results to be
real and with high statistical significance.

Interestingly, the same group, with the same methodology, obtained the opposite
trend for FG stars in NGC~3201. From the data of M19, shown in the right panel
of Fig.~\ref{f:feteff}, there seems to be a gradient, with cooler stars showing
slightly higher metallicities. However, the two-tail probability for the
correlation is $p=0.204$ (18 stars), so formally the correlation does not have
high statistical significance and it is likely to be only apparent, probably
driven only by the most metal-poor star, which is indicate as a candidate binary
star in M19. 

The treatment of candidate binary stars is another source of discrepancies
between the analyses of M23 and M19. In NGC~3201 the three probable binaries 
are all confined in the $RG1$ region at the most negative
values, and  are excluded by M19 from the correlations. Instead, in 47~Tuc the
two stars that are probable binary (according to their large rms values in
radial velocity) span the whole range of $\Delta col$ values  and are considered
in the analysis and ensuing correlations. Accordingly, we removed the 3 binary
stars in NGC~3201 from considerations and  kept the two stars in 47~Tuc. In
Fig.~\ref{f:feteff} (right panel) the Pearson's coefficient and the two-tail
probability are those referred to 15 stars, without the candidate binaries.

\begin{figure*}
\centering
\includegraphics[scale=0.30]{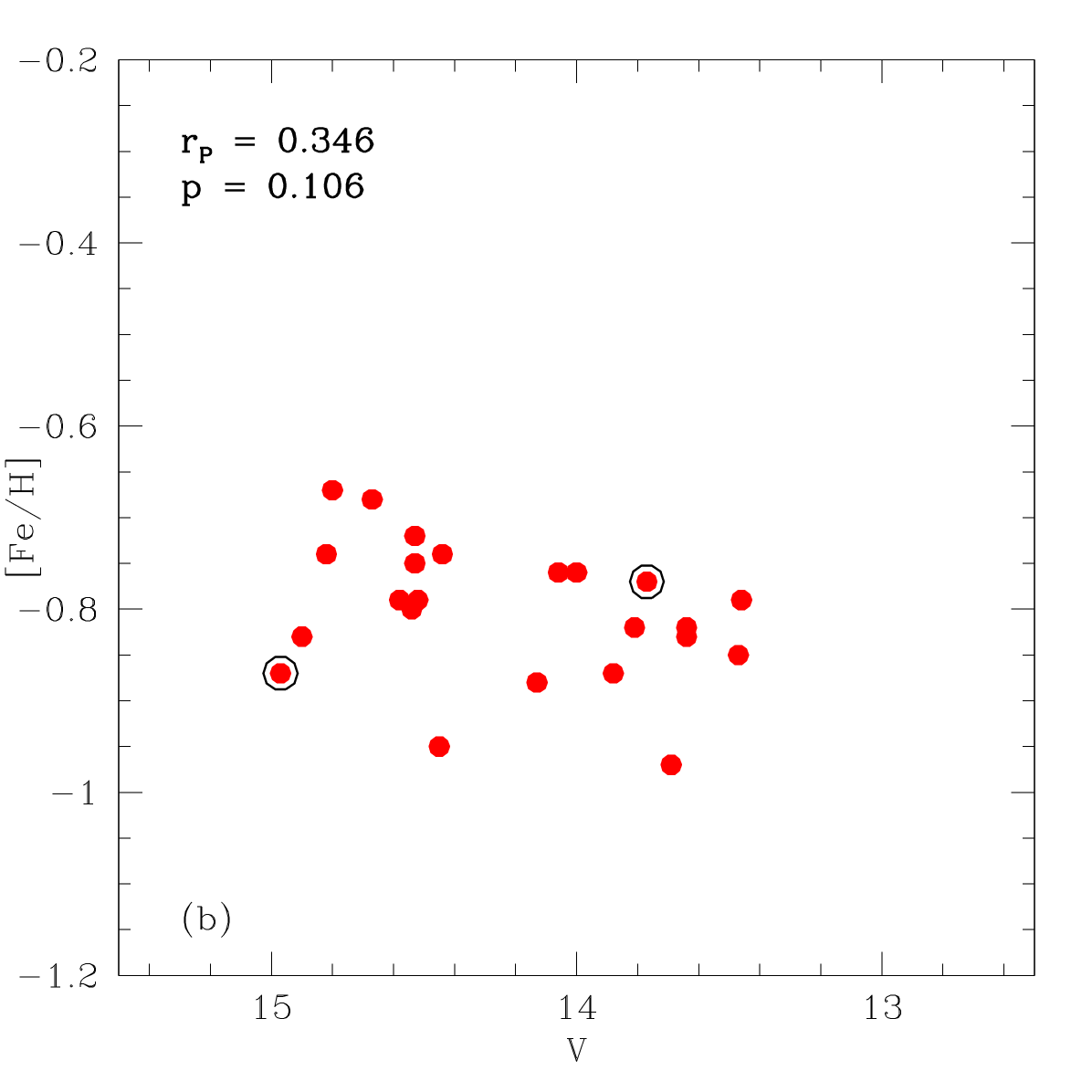}\includegraphics[scale=0.30]{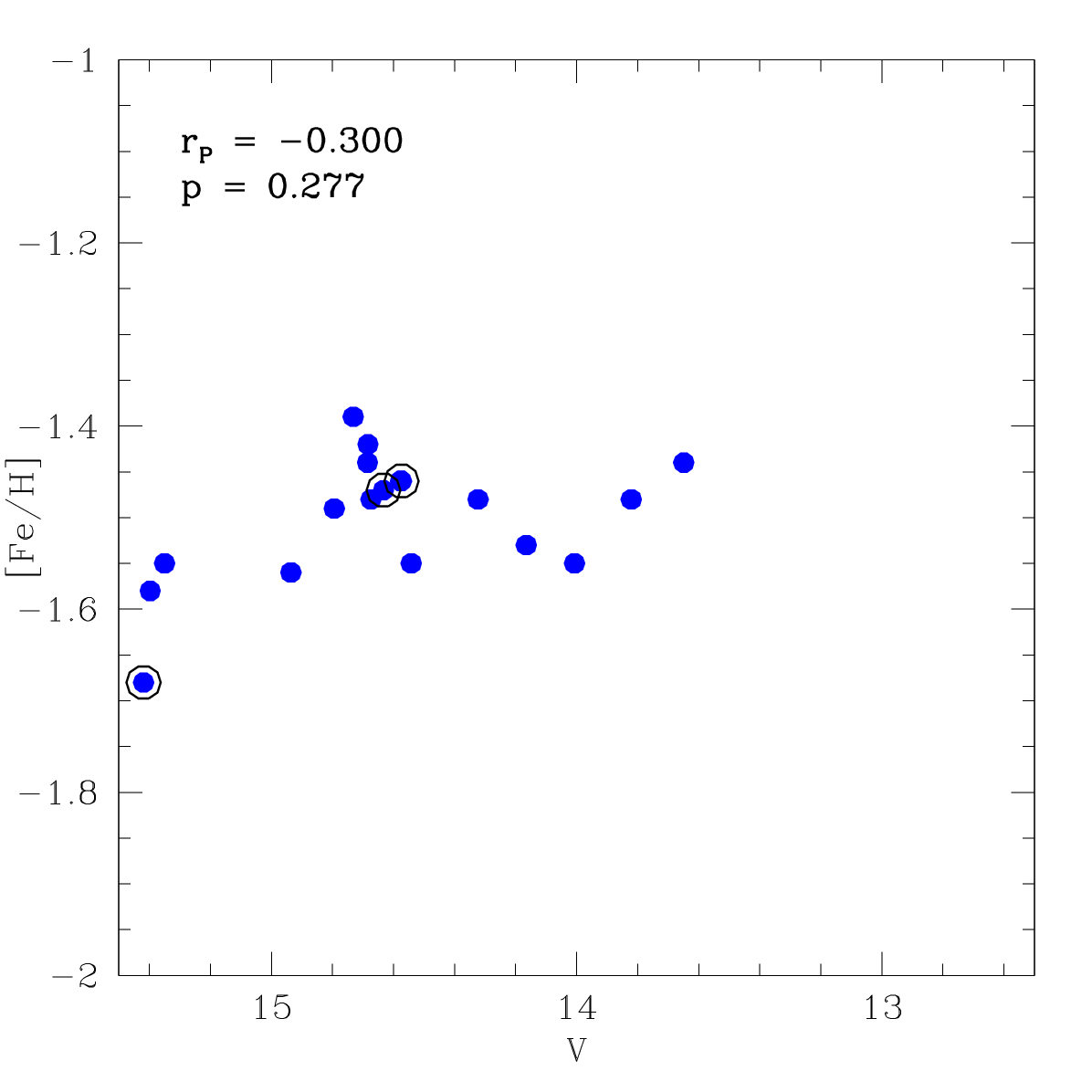}
\caption{Metallicity of FG stars in 47~Tuc (left panel) and in NGC~3201 (right
panel) as a function of the apparent $V$ magnitude, from M23 and M19,
respectively.}
\label{f:feV}
\end{figure*}

We examined the remote possibility that cooler stars in 47~Tuc were
fortuitously clustered at low metallicity, with all the metal-rich stars in the
sample having casually higher temperatures. To test whether this is the case we
show Fig.~\ref{f:feV} where the metallicities derived by M23
are plotted as a function of the $V$ magnitudes of stars, tabulated in M23.
Formally the observed trend (brighter stars having lower [Fe/H] values) does not
have high statistical significance, however it reflects the trend of
Fig.~\ref{f:feteff}. There is no astrophysical process able to generate such a
trend among RGB stars, so the implication is that this must be a spurious trend,
due to some problems in the spectroscopic analysis and the derived atmospheric
parameters.

Also in NGC~3201 the gradient of metallicities as a function of the $V$
magnitude does reflect the trend observed in Fig.~\ref{f:feteff}, although its
statistical significance is not high, in particular when neglecting the
candidate binaries.

We can check the claim about the relation between displacement in 
$\Delta col$ along $RG1$ and metallicity through the link with the effective
temperature, that on the RGB is strongly mediated by the metal abundance.

\begin{figure*}
\centering
\includegraphics[scale=0.30]{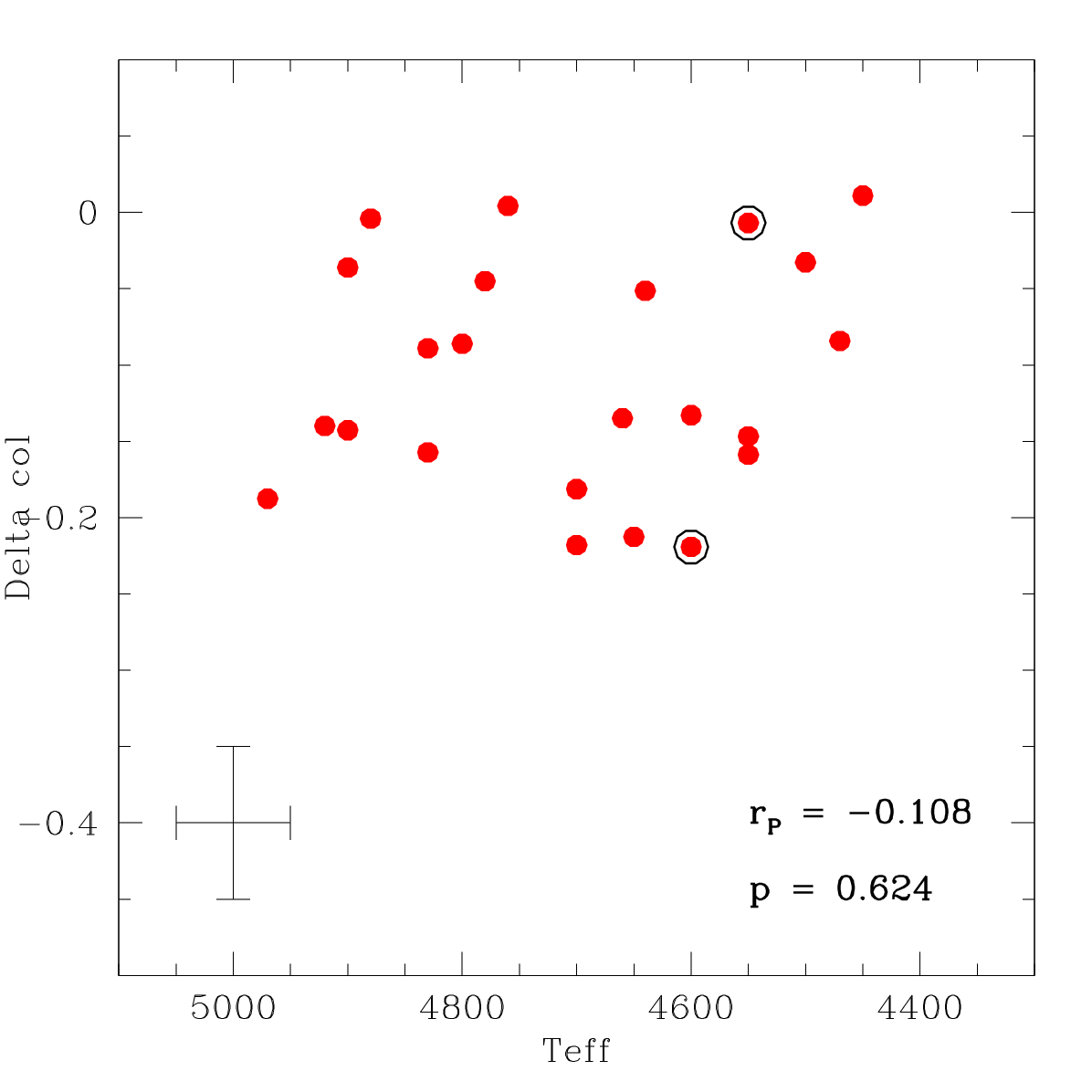}\includegraphics[scale=0.30]{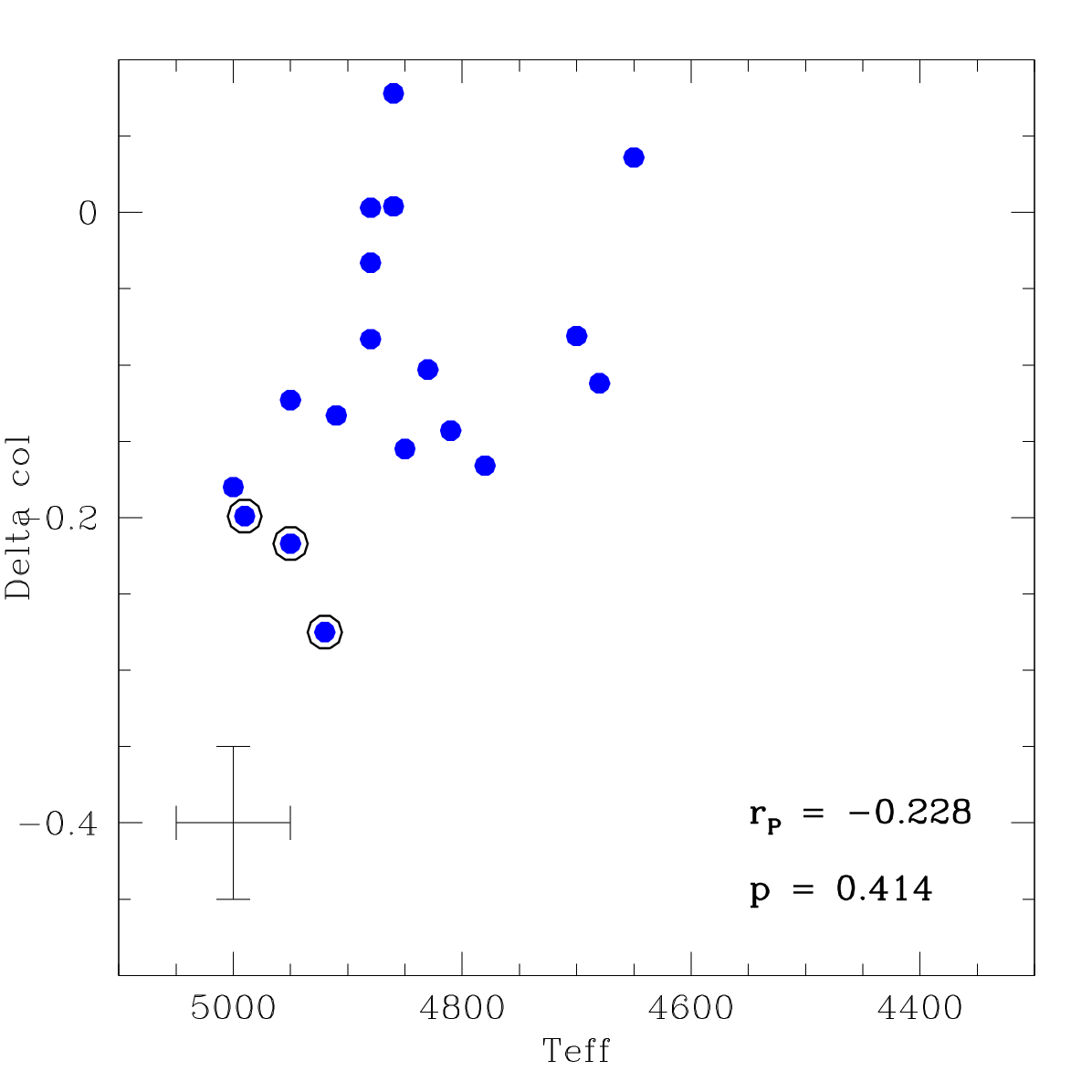}
\caption{Coordinate $\Delta col$ along the $RG1$ in the PCM of 47~Tuc as a
function of the effective temperature of stars from M23 (left panel). Right
panel: the same for NGC~3201 from M19. Typical error-bars from M19, M23, and
M17 are indicated.}
\label{f:teffdcol}
\end{figure*}

In Fig.~\ref{f:teffdcol} (left panel) we plotted the sample of FG stars in
47~Tuc with the values tabulated in M23. The run of $\Delta col$ as a function
of T$_{\rm eff}$ is clearly flat, there is no dependence on the temperature of
stars. Hence, there is no evidence for a variation of metallicity when moving
along this coordinate in the PCM locus populated by FG stars.

For NGC~3201 (right panel of Fig.~\ref{f:teffdcol}) there is a hint of a
trend, with stars at lower values in $\Delta col$ being slightly warmer, hence 
supposedly more metal-poor, as claimed. Nevertheless, the relation showed is 
formally not of high statistical significance, in particular excluding the
binaries, as done in M19.

\begin{figure*}
\centering
\includegraphics[scale=0.22]{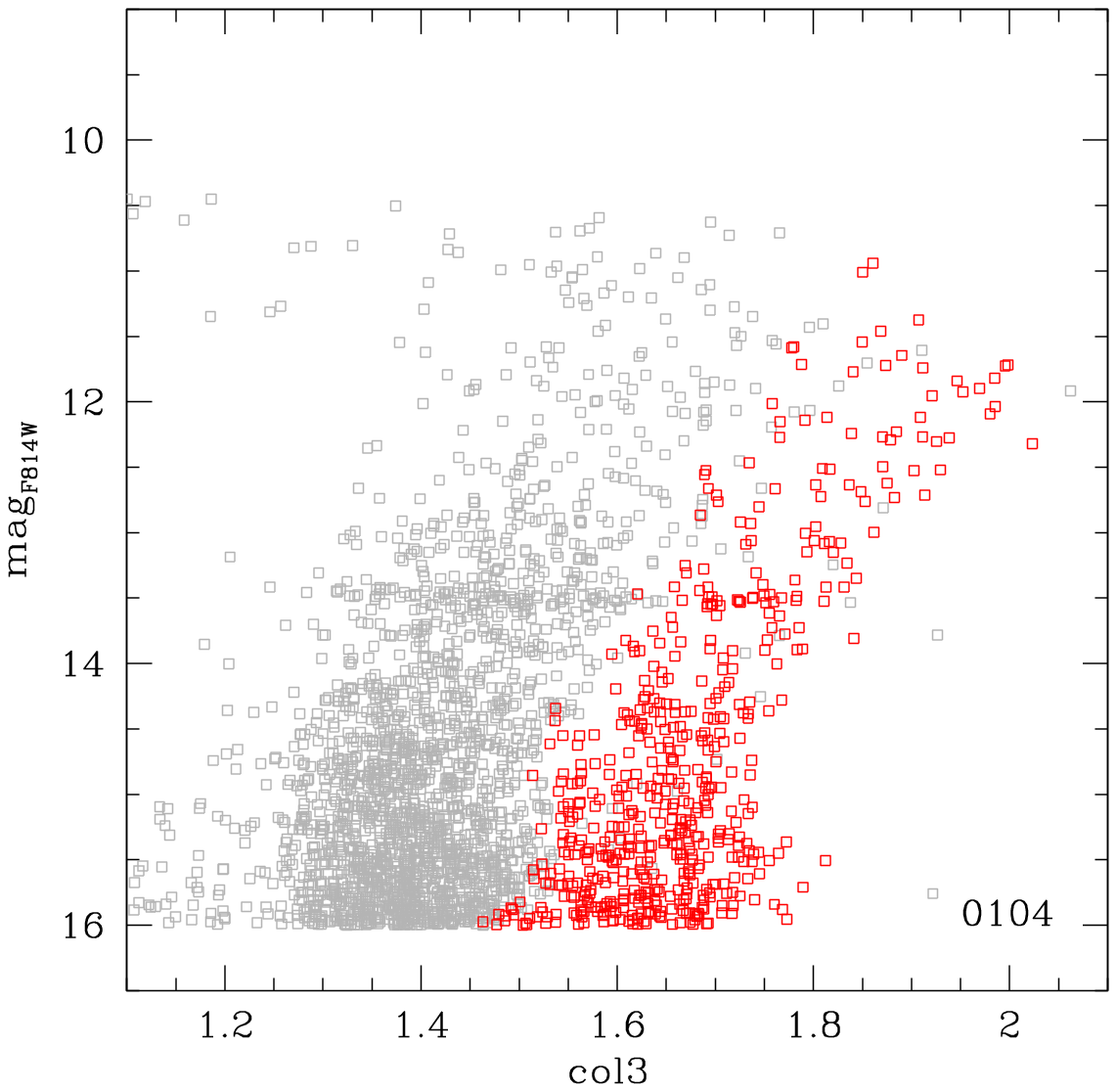}\includegraphics[scale=0.22]{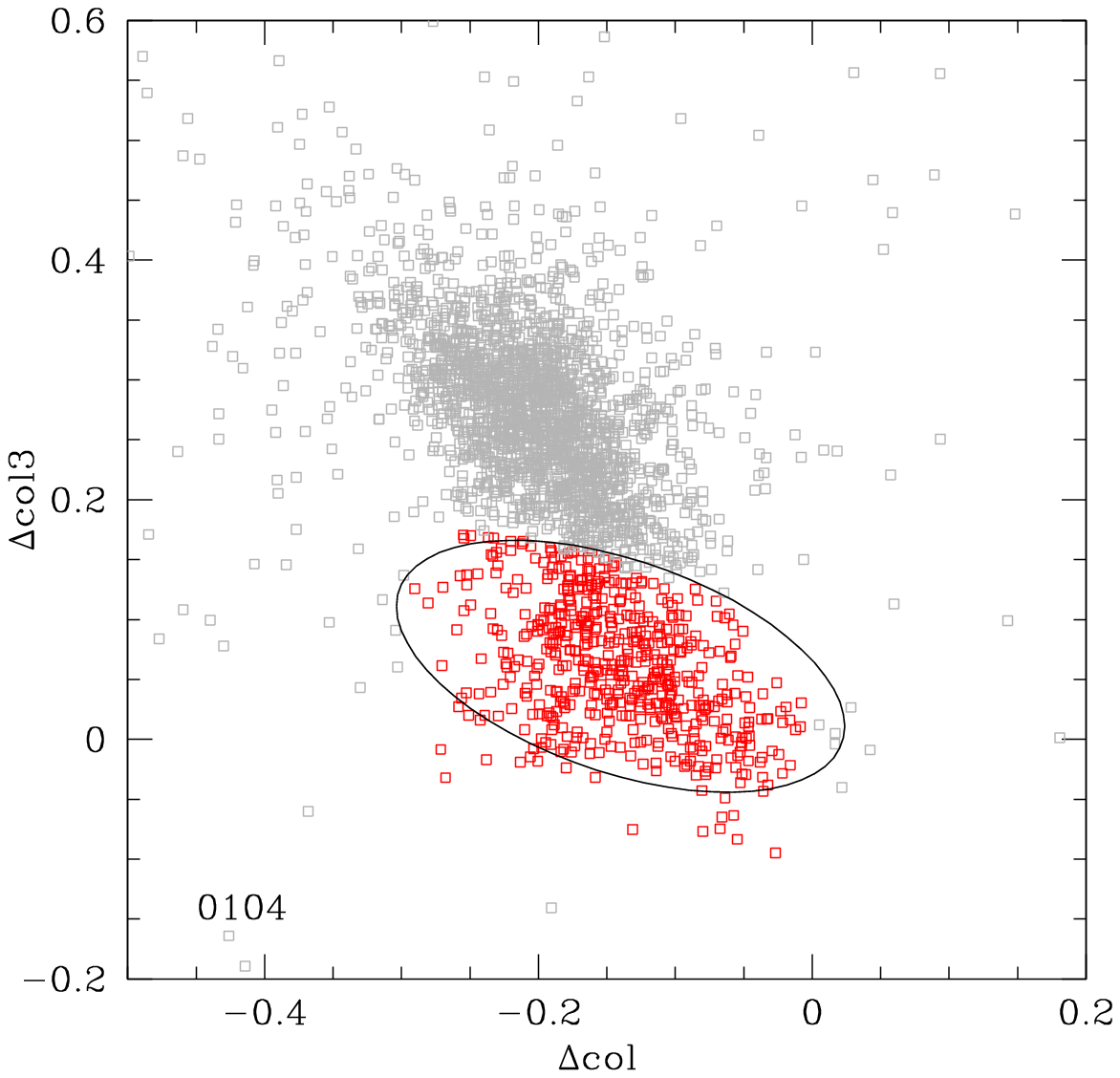}\includegraphics[scale=0.22]{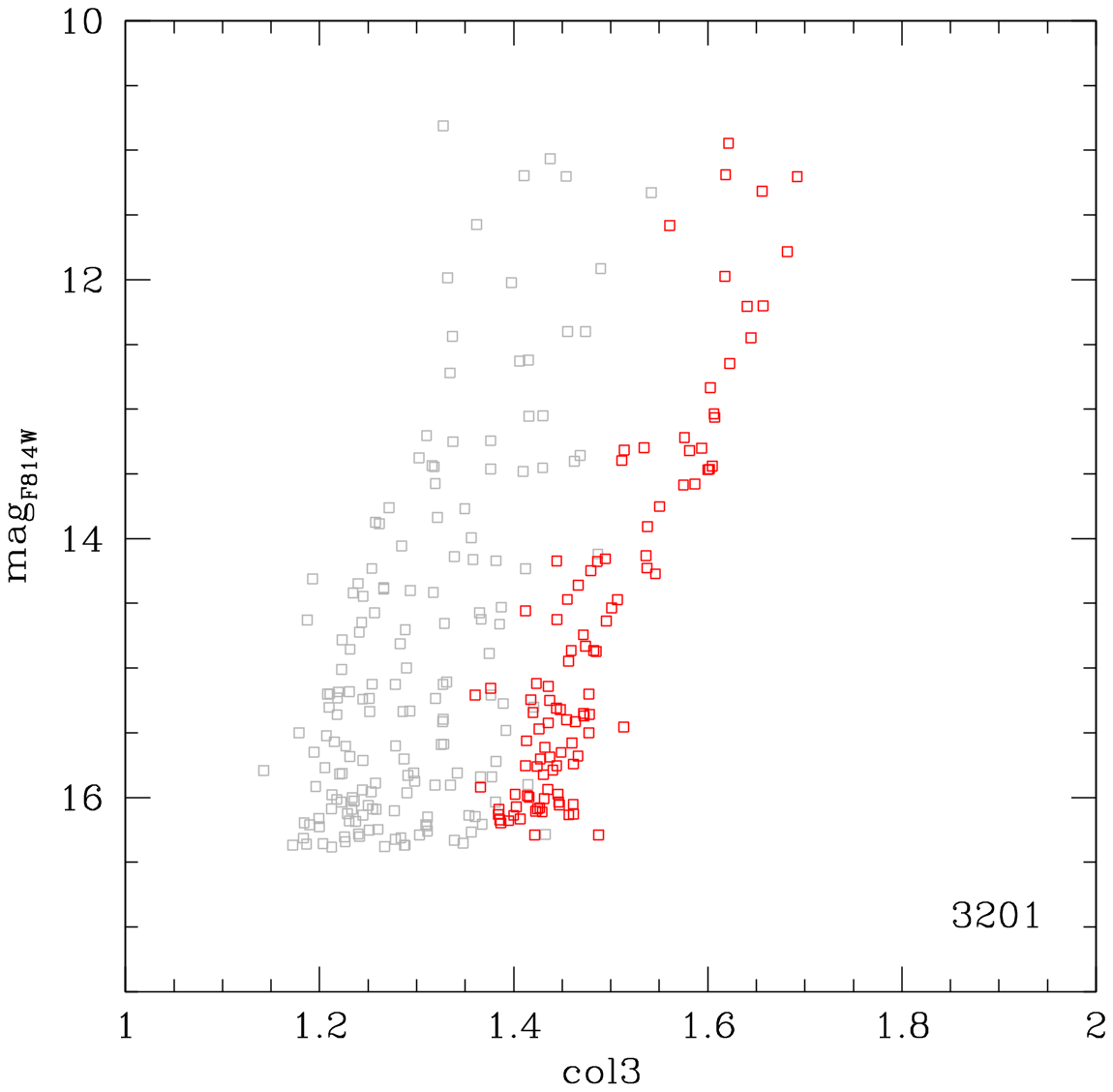}\includegraphics[scale=0.22]{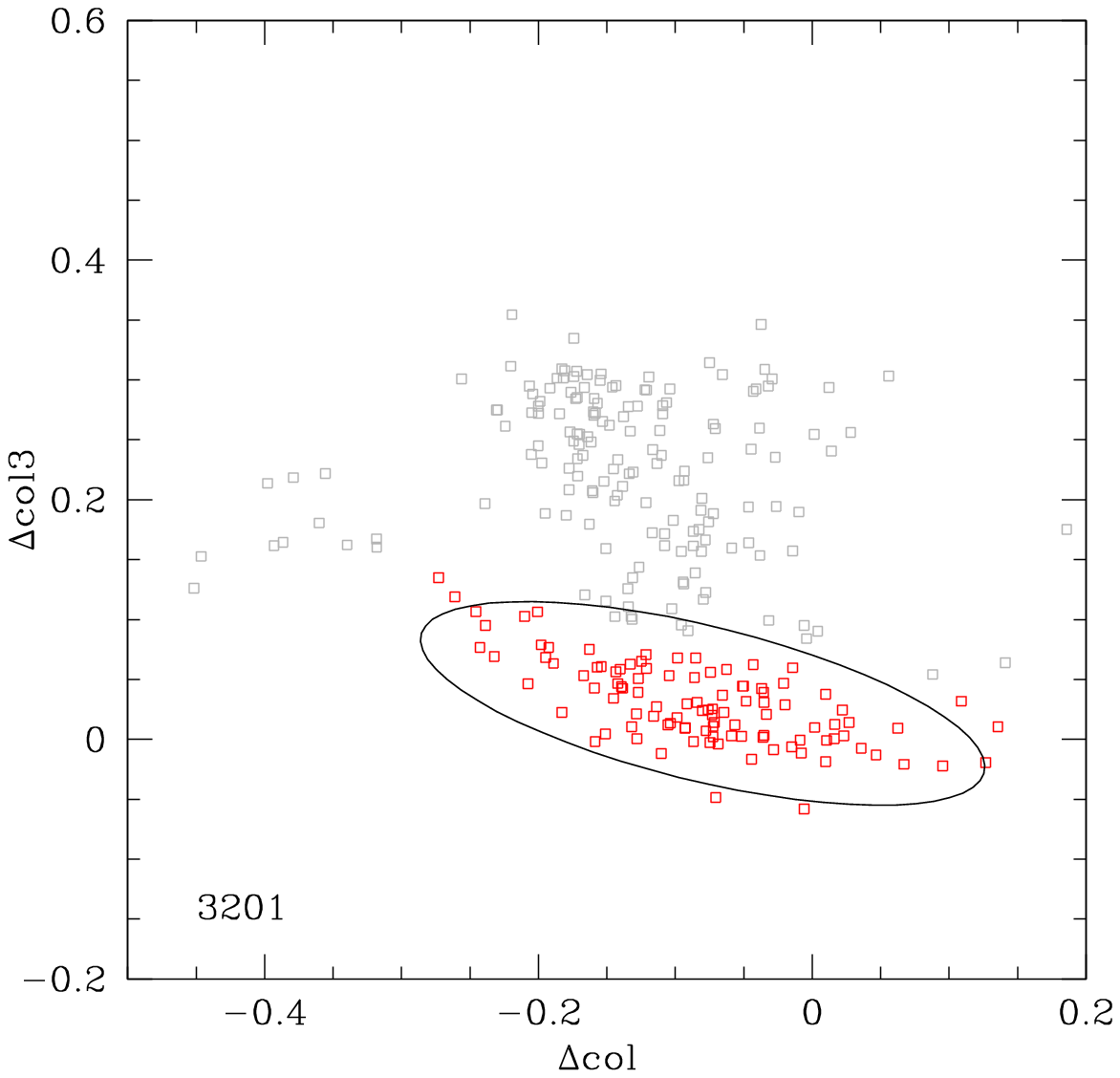}
\caption{Plots of $col3$ versus $m_{F814W}$ (left panels) and PCMs for 47~Tuc 
(upper row) and NGC~3201 (lower row), taken from Carretta and Bragaglia (2024).
In all panels FG stars are plotted in red. Ellipses enclosing FG stars in the
PCMs show the extension of $RG1$ in the two GCs.}
\label{f:ellissi}
\end{figure*}

In a forthcoming paper (Carretta and Bragaglia, in preparation) we
make an extensive exploration of the precise lengths of the $RG1$ regions in a
sample of 20 GCs. In Fig.~\ref{f:ellissi} we show a direct comparison of the
extent of $RG1$ in the two GCs under scrutiny in the present paper.

The extension of each $RG1$ locus is represented by ellipses enclosing the FG
stars selected in the left panel of Fig.~\ref{f:ellissi}. This figure allows to
confirm the result based on the much sparser data published ($\Delta col$ values
for 23 stars in 47~Tuc and 18 stars in NGC~3201 from M23 and M19, respectively).
The length of $RG1$ in $\Delta col$ is about 0.3 mag for 47~Tuc and 
0.5 mag in NGC~3201. Yet the claim by M23 and M19 is that the same metallicity
variation ($\sim 0.1$ dex) is apparently detected in both GCs. 

\section{Final considerations and summary}

In this paper we addressed the issue of possible metallicity variations among FG
stars. These stars are the outcome of the primary star formation phase in GCs
and are not supposed to be polluted by matter processed by proton-capture
reactions in H-burning at high  temperature. To test the claim that the
different extent of the $RG1$ in the PCM is related to metallicity spread we
compared the abundance analysis of FG stars in 47~Tuc and NGC~3201, performed by
the same group, with identical methodology, that is a fully spectroscopic
derivation of atmospheric parameters and metallicity [Fe/H] based on high
resolution UVES spectra.

In 47~Tuc we uncovered clear trends of the metallicity as a function of  T$_{\rm
eff}$ and luminosity among the 23 FG stars, with cooler (and brighter) stars
being more metal-poor than warmer (and fainter) giants along the RGB. These
gradients are unambiguous, cannot be explained by any known astrophysical
mechanism, and must be probably attributed to some problems in the abundance
analysis.

These problems bear consequences on the conclusion drawn from the abundance
analysis of FG stars in 47~Tuc. From the comparison of the abundance spread (or
lack thereof) in FG and SG stars, M23 advanced a scenario where FG stars
were born in a inhomogeneous medium during the first phase of the star 
formation, whereas the next generation formed more centrally concentrated in
regions where a cooling flow was able to more efficiently mix the gas used to
produce the polluted population.

\begin{figure}
\centering
\includegraphics[scale=0.42]{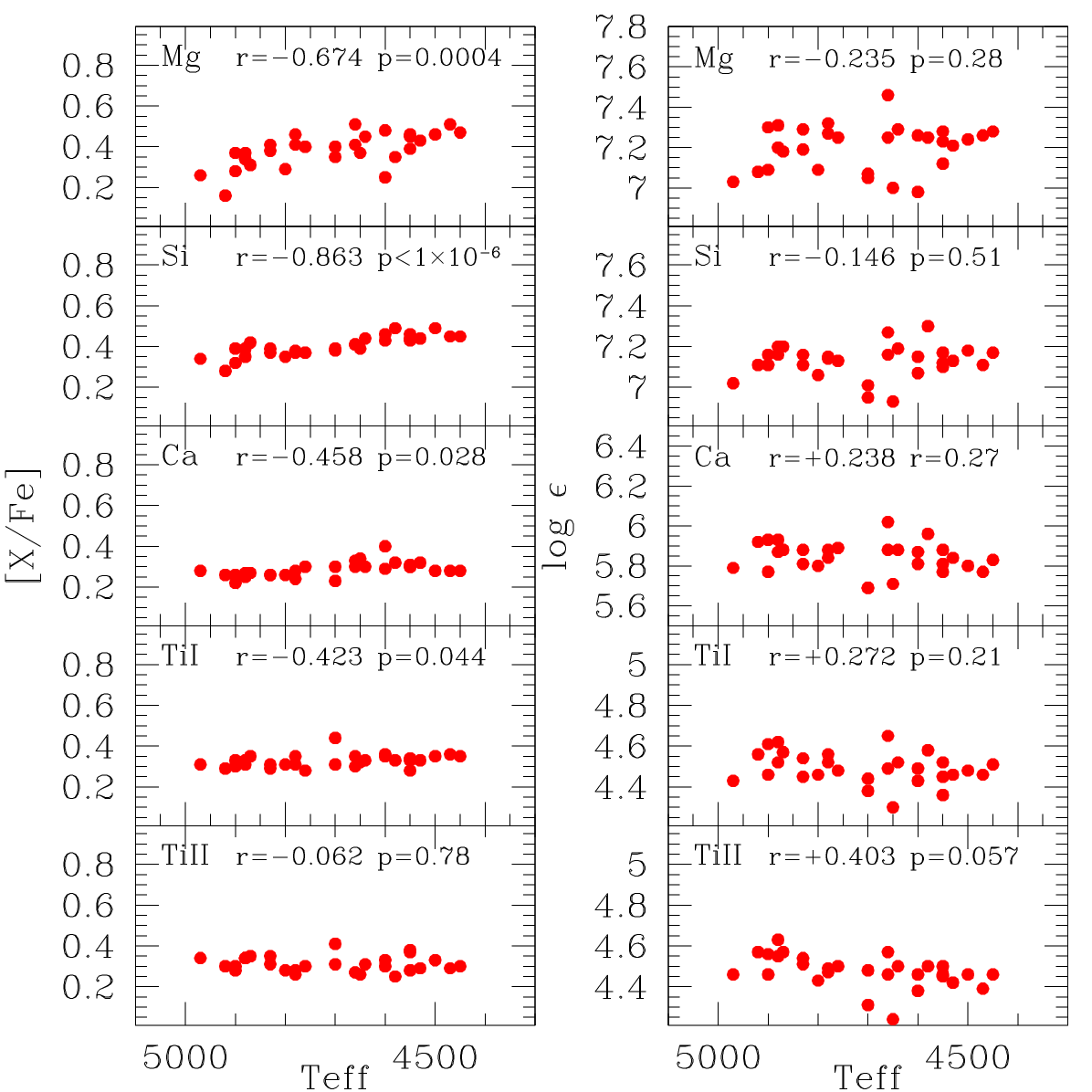}
\caption{Abundances of $\alpha-$elements as a function of T$_{\rm eff}$ in 23 FG
stars of 47~Tuc from M23. Left column: abundance ratios [el/Fe]. Right column:
absolute number density abundances. In each panel are reported the Pearson's
correlation coefficient and the two-tails probability of the linear regression with
T$_{\rm eff}$.}
\label{f:alpha}
\end{figure}

This scenario is however weakened by the spurious trend with effective
temperature uncovered for iron. The inhomogeneities in other species can be
simply the reflection of this trend. A countercheck is offered by
Fig.~\ref{f:alpha}. In this figure we plot the $\alpha-$elements Mg, Si, Ca,
Ti~{\sc i} and Ti~{\sc ii} as a function of the T$_{\rm eff}$ for all 23 FG stars in
47~Tuc from M23. In the panels of the left column, the abundance ratios
[X/Fe] are shown. The Pearson's correlation coefficient and the two-tail
probability of the linear regression are reported in each panel. All elements
except Ti~{\sc ii} show a correlation with T$_{\rm eff}$ with high statistical
significance. However, when we exclude the contribution of iron and we simply
plot the absolute abundance (in number density) of each species (panels in the
right column) the run is compatible with a constant level or, at least, absence
of any correlation with statistical significance. Furthermore, while the ratio
[Ti/Fe]~{\sc ii} was flat with T$_{\rm eff}$, the run of  
$\log \epsilon$(Ti~{\sc ii}) shows a noticeable trend that has almost a  
statistical significance.

\begin{figure}
\centering
\includegraphics[scale=0.42]{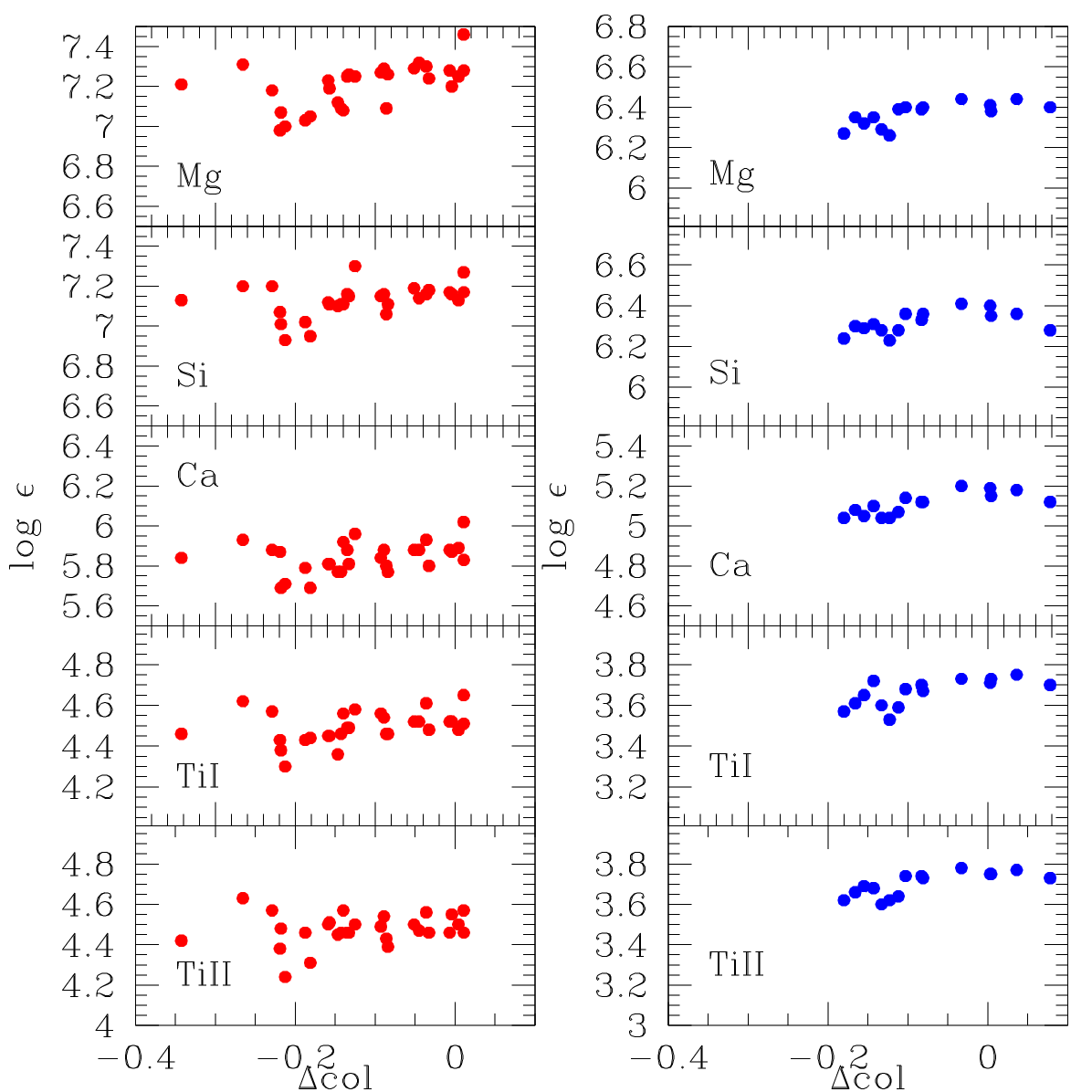}
\caption{Absolute number density abundances of $\alpha-$elements in FG stars of 
47~Tuc (left panels, from M23) and of NGC~3201 (right panels, from M19) as a 
function of the displacement $\Delta col$ on the $RG1$ region of the PCMs.}
\label{f:alphaadd}
\end{figure}

Hence, it is quite difficult to envisage a coherent scenario of a primordial
proto-cluster medium where iron is not well mixed, but other products of SN
nucleosynthesis like the $\alpha-$elements are homogeneous to such an extent 
not to show any significant variation with T$_{\rm eff}$, as implied in 
Fig.~\ref{f:alpha}.

Proofs that SG stars formed generally in denser and likely more well mixed 
medium do exist. The lower binary fraction among SG stars strongly suggests
different conditions of the environments where unpolluted and polluted stars
formed. High efficiency of infant mortality of binary systems among SG stars
nicely explains the observed difference in binary fractions (D'Orazi et al. 2010,
Lucatello et al. 2015). On the other hand, data shown in Fig.~\ref{f:alpha} do
not seem adequate enough to confirm such a scenario.

The situation is less uncertain for NGC~3201. With the notable exceptions of Mg,
Cu, and Pr, no species in NGC~3201 show a trend with T$_{\rm eff}$ (in both
abundance ratio with iron and absolute elemental abundance) which has a
statistical significance\footnote{Moreover, the trend in Pr abundances,
decreasing as T$_{\rm eff}$ decreases, is opposite to the ones for Mg and Cu.}.
The absence of spurious trends derives from the nearly constant values in
Fig.~\ref{f:feteff}, when candidate binaries are excluded.

In both GCs there seems to be no compelling evidence for a variation in 
T$_{\rm eff}$ for the FG samples as a function of the $\Delta col$ coordinate.
Since the T$_{\rm eff}$ of giants on the RGB is a strong function of the
metallicity of stars, overall these findings cast some doubts on the claim that
there is a metallicity spread among FG stars, or at least weakens its
relevance.

On the other hand, we confirm that in both GCs the absolute chemical abundances
for most of the species analysed present a trend to be higher in FG stars
with $\Delta col$ near zero, and lower when going to $\Delta col$ increasingly
negative values (see Fig.~\ref{f:alphaadd} for an example of the trends
with $\Delta col$ of the $\alpha-$elements in 47~Tuc and NGC~3201). The few
exceptions are for Al, Sc, Co, Y, Zr, Pr in NGC~3201 and for Zn, Y, Ba, La in
47~Tuc. This could be surprising, since $col$ and  $\Delta col$ should not be
influenced by the abundances of these elements (with  the possible exception of
oxygen). The variation, essentially due to temperature, should be related only
to changes in helium, or metallicity. Considering the lack of significant
variations of T$_{\rm eff}$ with $\Delta col$ it is not clear how different
abundances can be associated to different positions along the horizontal
coordinate in the PCMs.

Finally, we stress that both analyses (identical as far as methodology is
concerned) compared in the present paper are deriving the same metallicity 
spread in GCs with clearly different extension of the $RG1$ region of FG stars
and global metallicity [Fe/H], despite the abundance analysis in 47~Tuc
producing spurious trends incompatible with stellar physics. The problem of the
possible existence of metallicity spreads among FG stars in multiple populations
of GCs seems still  to remain an open issue at the moment.

\begin{acknowledgements}
We thank the anonymous referee for their fair and constructive comments.
This research has made use of the VizieR catalogue access tool, CDS, 
 Strasbourg, France (DOI: 10.26093/cds/vizier). The original description of the
 VizieR service was published in 2000, A\&AS 143, 23. Use of the NASA's
Astrophysical Data System, and TOPCAT (Taylor 2005) are also acknowledged.
\end{acknowledgements}


\begin{thebibliography}{}

\bibitem[]{} Bastian, N., Lardo, C. 2018, ARA\&A, 56, 83
\bibitem[]{} Bekki, K., Tsujimoto, T. 2016, ApJ, 831, 70
\bibitem[]{} B\"ohm-Vitense, E. 1979, ApJ, 234, 521
\bibitem[]{} Carretta, E. 2015, ApJ, 810, 148 
\bibitem[]{} Carretta, E., Bragaglia, A., Gratton R.G., et al. 2006, A\&A, 450, 523 
\bibitem[]{} Carretta, E., Bragaglia, A., Gratton, R.G., D'Orazi, V., Lucatello,
 S. 2009a, A\&A, 508, 695 
\bibitem[]{} Carretta, E., Bragaglia, A., Gratton, R.G. et al. 2010a, ApJ, 714, L7 
\bibitem[]{} Carretta, E., Lucatello, S., Gratton, R.G., Bragaglia, A., D'Orazi,
  V. 2011, A\&A, 533, 69 
\bibitem[]{} D'Orazi, V., Gratton, R., Lucatello, S. et al. 2010, ApJ, 719, L213
\bibitem[]{} Gratton, R.G., Sneden, C., Carretta, E. 2004, ARA\&A, 42, 385
\bibitem[]{} Gratton, R.G., Carretta, E., Bragaglia, A. 2012, A\&ARv, 20, 50 
\bibitem[]{} Gratton, R.G., Bragaglia, A., Carretta, E., et al. 2019, A\&ARv, 27, 8
\bibitem[]{} Decressin, T., Meynet, G., Charbonnel C. Prantzos, N.,
 Ekstrom, S. 2007, A\&A, 464, 1029 
\bibitem[]{} de Mink, S.E., Pols, O.R., Langer, N., Izzard, R.G. 2009, A\&A,
  507, L1 
\bibitem[]{} Denissenkov, P.A., Hartwick, F.D.A. 2014, MNRAS, 437, L21 
\bibitem[]{} Johnson, C.I., Pilachowski, C.A. 2010, ApJ, 722, 1373 
\bibitem[]{} Lardo, C., Salaris, M., Cassisi, et al. 2023, A\&A, 669, A19
\bibitem[]{} Legnardi, M.V., Milone, A.P., Armillotta, L. et al. 2022, MNRAS,
  513, 735
\bibitem[]{} Lucatello, S., Sollima, A., Gratton, R., et al. 2015, A\&A, 584,
  A52
\bibitem[]{} Marino, A.F., Milone, A.P., Sills, A. et al. 2019, ApJ, 887, 91
\bibitem[]{} Marino, A.F., Milone, A.P., Dondoglio, E. et al. 2023, ApJ, 958, 31
\bibitem[]{} Milone, A.P., Piotto, G., Renzini, A. et al. 2017, MNRAS, 464,
  3636 
\bibitem[]{} Mucciarelli, A., Bellazzini, M., Ibata, R., et al. 2012, MNRAS, 426, 2889 
\bibitem[]{} Salaris, M., Cassisi, S., Weis, A. 2002, PASP, 114, 375
\bibitem[]{} Taylor, M.B. 2005, Astronomical Data Analysis Software and Systems
  XIV, 347, 29
\bibitem[]{} Tsujimoto, T., Shigeyama, T. 2003, ApJ, 590, 803
\bibitem[]{} Ventura, P. D'Antona, F., Mazzitelli, I., Gratton, R. 2001,
  ApJ, 550, L65 

\end{thebibliography}
\end{document}